\documentclass[11pt]{article}
\def\lb{\label}
\def\be{\begin{equation}}
\def\ee{\end{equation}}
\def\ba{\begin{eqnarray}}
\def\ea{\end{eqnarray}}

\def\bb{\bibitem}

\def\e{{\rm e}}
\begin{document}
\title{
\begin{flushright}\begin{small}
LAPTH-1178/07
\end{small}\end{flushright}
\vspace{2cm}
Hawking radiation of linear dilaton black holes}
\author{
G. Cl\'ement$^{a}$\footnote{e-mail: gclement@lapp.in2p3.fr},
J.C. Fabris$^{b}$\footnote{e-mail: fabris@cce.ufes.br}
and G.T. Marques$^{a,b}$\footnote{e-mail:gtadaiesky@cce.ufes.br}\\ \\
{\small $^a$Laboratoire de  Physique Th\'eorique LAPTH (CNRS),} \\
{\small B.P.110, F-74941 Annecy-le-Vieux cedex, France}\\ {\small
$^{b}$ Departamento de F\'{\i}sica, Universidade Federal do
Esp\'{\i}rito Santo,}\\ {\small Vit\'oria, 29060-900,
Esp\'{\i}rito Santo, Brazil}}
\date{April 3, 2007}
\maketitle

\begin{abstract}
We compute exactly the semi-classical radiation spectrum for a class of
non-asymptotically flat charged dilaton black holes, the so-called
linear dilaton black holes. In the high frequency regime, the temperature
for these black holes generically agrees with the surface gravity result.
In the special case where the black hole is massless, we show
that, although the surface gravity remains finite, there is no radiation,
in agreement with the fact that massless objects cannot radiate.
\end{abstract}
\newpage

Quantum field theory in curved spacetime predicts new phenomena such
as particle emission by a black hole \cite{birrell}. This is due to
the fact that the vacuum for a quantum field near the horizon is
different from the observer's vacuum at spatial infinity. A distant
observer thus receives from a black hole a steady flux of particles
exhibiting, in the high frequency regime, a black body spectrum with
a temperature proportional to the surface gravity \cite{haw}.
Although Hawking's original derivation of this black hole
evaporation dealt with realistic collapsing black holes, Unruh
\cite{unruh1} showed that the same results are obtained when the
collapse is replaced by appropriate boundary conditions on the
horizon of an eternal black hole. In the semi-classical
approximation, the black hole radiation spectrum may be evaluated by
computing the Bogoliubov coefficients relating the two vacua. An
equivalent procedure is to compute the reflection and absorption
coefficients of a wave by the black hole. Usually, the wave equation
cannot be solved exactly, and one must resort to match solutions
in an overlap region between the near-horizon and asymptotic regions
\cite{page,unruh2}. In the special case of the (2+1)-dimensional BTZ
black hole \cite{BTZ}, an exact solution of the wave equation is
available, which allows for an exact computation of the radiation
spectrum, leading to the Hawking temperature \cite{GL,NSS,BSS}.

In this Letter, we discuss another case of black holes also allowing
for an exact semi-classical computation of their radiation spectrum,
that of linear dilaton black hole solutions to Einstein-Maxwell
dilaton (EMD) theory in four dimensions. Linear dilaton black holes
are a special case of the more general class of non-asymptotically
flat black hole solutions to EMD \cite{CHM,newdil2}, which we first
briefly present. We discuss the evaporation of these
non-asymptotically flat black holes and show that they either
collapse to a naked singularity in a finite time, or evaporate in an
infinite time. We then specialize to linear dilaton black holes, and
outline the analytical computation of their radiation spectrum. For
massive black holes, this computation leads, in the high frequency
regime, to the same temperature which is obtained from the surface
gravity. However in the case of massless extreme black holes, we find
that, although the surface gravity remains finite, there is no
radiation, in agreement with the fact that a massless object cannot
radiate.

EMD is defined by the following action
\begin{equation}
\label{lagran}
S = \frac{1}{16\pi}\int dx^4\sqrt{-g}\biggr[R -
2\partial_\mu\phi\partial^\mu\phi -
e^{-2\alpha\phi}F_{\mu\nu}F^{\mu\nu}\biggl] \quad ,
\end{equation}
where $F_{\mu\nu}$ is the electromagnetic field, and $\phi$
is the dilatonic field, with coupling constant $\alpha$. This theory
admits static spherically symmetric solutions representing black
holes. Among these black hole solutions there are asymptotically
flat ones \cite{GM,GHS} as well as non-asymptotically flat
configurations \cite{CHM,newdil2}. In the present work, we are
interested in the non-asymptotically flat black hole solutions
\ba
ds^2 &=& \frac{r^\gamma(r - b)}{r_0^{\gamma + 1}}dt^2 -
\frac{r_0^{\gamma + 1}}{r^\gamma(r - b)}\biggr\{dr^2 + r(r -
b)d\Omega^2\biggl\}\,, \lb{nbhm}
\\ F & = &
\sqrt{\frac{1+\gamma}2}\,\frac{\nu}{r_0}\,dr\wedge dt\,, \qquad
e^{2\alpha\phi} = \nu^{2}\bigg(\frac{r}{r_0}\bigg)^{1-\gamma}\,. \lb{nbhe}
\ea
with
\begin{equation}
\gamma = \frac{1 - \alpha^2}{1 + \alpha^2}\,.
\end{equation}
The constants $b$ and $r_0$ are related to the mass and to the
electric charge of the black hole through
\be
M = (1 - \gamma)b/4\,, \qquad Q =
\sqrt{\frac{1+\gamma}2}\frac{r_0}{\nu}\,.
\ee

The solutions (\ref{nbhm}),(\ref{nbhe}) interpolate between the
Schwarzschild solution for $\gamma = -1$ ($\alpha^2 \to \infty$) and
the Bertotti-Robinson solution for $\gamma = +1$ ($\alpha^2 = 0$). For
$b > 0$ the horizon at $r = b$ hides the singularity at $r = 0$, while
in the extreme black hole case $b = 0$ the horizon coincides with
the singularity. This is a curious case, with vanishing mass but a
finite electric charge. For $-1< \gamma < 0$ ($\alpha^2 > 1$) the
central singularity is timelike and clearly naked \cite{newdil2}. On
the other hand, for $0 \le \gamma < 1$ ($0 < \alpha^2 \le 1$),
the central singularity is null
and marginally trapped \cite{hayward}, so that signals coming from
the centre never reach external observers. Thus in this case, extreme
black holes can be still considered as black holes indeed.

The statistical Hawking temperature of the black holes (\ref{nbhm}),
computed as usual by dividing the surface gravity by $2\pi$ is given
by
\begin{equation}\lb{th}
T_H = \frac{1}{4\pi}\frac{b^\gamma}{r_0^{1+\gamma}} \,.
\end{equation}
It is finite for all $\gamma$ if $b \neq 0$. For $b = 0$ and $-1 <
\gamma < 0$ (naked singularity). the temperature is infinite, while
for $b = 0$ and $0 < \gamma < 1$ (extreme black hole), the temperature
vanishes.

The case $b = \gamma = 0$  is intriguing. Although this an extreme
black hole, the situation is different from that of
asymptotically flat extreme black holes. The near-horizon Euclidean extreme
Reissner-Nordstr\"om geometry is cylindrical, rather than conical, so
that its statistical temperature is arbitrary, contrary to the zero
value derived from surface gravity \cite{HHR}. In the present case the
two-dimensional Euclidean continuation of the metric (\ref{nbhm}) with
$\gamma=0$ clearly has a conical singularity at $r=b$ for all values
of $b$, including $b=0$, leading for this particular extreme black
hole to the finite temperature $T_H = 1/4\pi r_0$, in agreement with
the value (\ref{th}). However this result is
questionable. A black hole with pointlike horizon and zero mass
clearly cannot radiate, so one should rather expect its temperature to
be zero. We will return to this question presently.

As black holes (\ref{nbhm}) radiate, they loose mass according to
Stefan's law
\be\lb{stef}
\frac{dM}{dt} = -\sigma A_hT_H^4\,,
\ee
where $\sigma$ is Stefan's constant, and
$A_h=4\pi r_0^{1+\gamma}b^{1-\gamma}$ is the horizon area. Assuming
that only electrically neutral quanta are radiated, (\ref{stef})
implies that the horizon area decreases according to
\be
\frac{db}{dt} =
-\frac{4\sigma}{(4\pi)^3(1-\gamma)}r_0^{-3(1+\gamma)}b^{1+3\gamma}\,,
\ee
which is solved by
\ba
b(t) &=& r_0\bigg(\frac{\gamma c}{1-\gamma}\,
\frac{t-t_0}{r_0^3}\bigg)^{-1/3\gamma}\qquad (\gamma \neq 0)\,,
\nonumber\\
b(t) &=& r_0\exp\bigg(-\frac{c}{3}\,
\frac{t-t_0}{r_0^3}\bigg) \qquad (\gamma = 0)\,,
\ea
where $c= 3\sigma/16\pi^3$, and $t_0$ is an integration constant.
The outcome depends on the
sign of $\gamma$. For $\gamma<0$, the Hawking temperature increases
with decreasing mass and the black hole collapses to a naked
singularity (or evaporates away altogether in the Schwarzschild case
$\gamma = -1$) in a finite time according to $b \sim
(t_0-t)^{1/3|\gamma|}$. On the other hand, for $\gamma \ge 0$, the
Hawking temperature decreases (or is constant for $\gamma=0$) with
decreasing mass, and the black hole evaporates in an infinite time,
reaching the extreme black hole state $b=0$ only asymptotically.

We now proceed to a more precise evaluation of the temperature of
non-asymptotically flat black holes from the study of wave
scattering in these spacetimes. The wave equation
\begin{equation} \label{kg} \nabla^2\phi = 0
\end{equation}
does not generically allow for an exact solution in the spacetimes
(\ref{nbhm}). However, it can be solved analytically \cite{newdil1}
in the case of linear dilaton black holes with $\gamma=0$ and $b\neq0$,
with the metric
\be\lb{nbhm0} ds^2 = \frac{r - b}{r_0}dt^2 -
\frac{r_0}{r - b}\biggr\{dr^2 + r(r - b)d\Omega^2\biggl\}\,, \ee

Considering the harmonic eigenmodes
\be
\phi(x) = \psi(r,t)Y_{lm}(\theta,\varphi)\,, \quad \psi(r,t) =
R(r)\e^{-i\omega t}\,,
\ee
we obtain the following radial equation:
\begin{equation}
\label{kgrad}
\partial_r\bigg(r(r - b)\partial_rR\bigg) +
\biggr(\bar\omega^2\frac{r}{r-b} - l(l + 1)\bigg)R = 0
\end{equation}
($\bar\omega^2 \equiv \omega^2 r_0^2$).
Putting
\begin{equation}
y = \frac{b-r}{b}\,, \quad R = y^{i\bar\omega}f\,,
\end{equation}
reduces (\ref{kgrad}) to the equation
\begin{equation}\lb{hyp1}
y(1-y)\partial^2_y f + \bigg(1+2i\bar\omega-2(1+i\bar\omega)y\bigg)
\partial_y f + \bigg(\bar\omega^2 - i\bar\omega - \bar\lambda^2
- 1/4\bigg)f = 0 \,,
\end{equation}
with
\begin{equation}
\bar\lambda^2 = \bar\omega^2 - (l + 1/2)^2 \,.
\end{equation}
This is a hypergeometric equation
\be
y(1-y)\partial^2_y f + \bigg(c-(a+b+1)y\bigg)\partial_y f -abf = 0\,,
\ee
with
\be
a = \frac12 + i(\bar\omega+\bar\lambda)\,, \quad b = \frac12 +
i(\bar\omega-\bar\lambda)\,, \quad c = 1 + 2i\bar\omega\,.
\ee
It follows that the general solution of equation (\ref{kgrad}) is
\begin{eqnarray}
&R& = C_1\bigg(\frac{r-b}{b}\bigg)^{i\bar\omega} F\bigg(\frac12 +
i(\bar\omega+\bar\lambda), \frac12 + i(\bar\omega-\bar\lambda),1+2i\bar\omega;
\frac{b-r}{b}\bigg) \nonumber\\
&+& C_2\bigg(\frac{r-b}{b}\bigg)^{-i\bar\omega} F\bigg(\frac12 -
i(\bar\omega+\bar\lambda), \frac12 - i(\bar\omega-\bar\lambda),1-2i\bar\omega;
\frac{b-r}{b}\bigg) \,.
\end{eqnarray}
Putting
\be\frac{r-b}b = \e^{x/r_0}\,,\ee the partial wave near the horizon
($x \to - \infty$) is thus
\be
\psi \simeq C_1 \e^{i\omega(x-t)} + C_2 \e^{-i\omega(x+t)}\,.
\ee

To obtain the behavior of the partial wave near spatial infinity, we
must expand the solutions of (\ref{hyp1}) in hypergeometric functions
of argument $1/y$. The relevant transformation is
\ba
F(a,b,c;y) &=&
\frac{\Gamma(c)\Gamma(b-a)}{\Gamma(b)\Gamma(c-a)}(-y)^{-a}
F(a,a+1-c,a+1-b;1/y) \nonumber\\
&+& \frac{\Gamma(c)\Gamma(a-b)}{\Gamma(a)\Gamma(c-b)}(-y)^{-b}
F(b,b+1-c,b+1-a;1/y)\,.
\ea
This leads to the asymptotic behavior
\be
\psi \simeq \bigg(\frac{r}b\bigg)^{-1/2}\bigg(B_1 \e^{i(\lambda x-\omega
  t)} + B_2 \e^{-i(\lambda x + \omega t)}\bigg)
\ee
($\lambda = \bar\lambda/r_0$), where the amplitudes of the asymptotic
outgoing and ingoing waves
$B_1$ and $B_2$ are related to the amplitudes of the near-horizon
outgoing and ingoing waves $C_1$ and $C_2$ by
\ba
B_1 &=& \Gamma(2i\bar\lambda)\bigg[\frac{\Gamma(1+2i\bar\omega)}
{\Gamma(1/2 + i(\bar\omega+\bar\lambda))^2}C_1
+ \frac{\Gamma(1-2i\bar\omega)}
{\Gamma(1/2 - i(\bar\omega-\bar\lambda))^2}C_2\bigg] \,, \nonumber\\
B_2 &=& \Gamma(-2i\bar\lambda)\bigg[\frac{\Gamma(1+2i\bar\omega)}
{\Gamma(1/2 + i(\bar\omega-\bar\lambda))^2}C_1
+ \frac{\Gamma(1-2i\bar\omega)}
{\Gamma(1/2 - i(\bar\omega+\bar\lambda))^2}C_2\bigg] \,.
\ea

Hawking radiation can be considered as the inverse process of
scattering by the black hole, with the asymptotic boundary condition
$B_1 = 0$ (the outgoing mode is absent). The coefficient for
reflection by the black hole is then given by
\be\lb{refl}
R = \frac{\vert C_1 \vert^2}{\vert C_2 \vert^2}\bigg\vert_{B_1=0}
= \frac{\vert\Gamma(1/2 + i(\bar\omega+\bar\lambda))^2\vert^2}
{\vert\Gamma(1/2 + i(\bar\omega-\bar\lambda))^2\vert^2}
=
\frac{\cosh^2\pi(\bar\omega-\bar\lambda)}
{\cosh^2\pi(\bar\omega+\bar\lambda)}\,.
\ee
The resulting radiation spectrum is
\be
N = \frac{R}{1-R} = (\e^{\omega/T_H}-1)^{-1}\,.
\ee
For high frequencies, $\bar\lambda \simeq \bar\omega = \omega/r_0$, and we
recover from
(\ref{refl}) the Hawking temperature as computed from the surface gravity,
\be
T_H = \frac1{4\pi r_0}\,.
\ee

The above computation fails in the linear dilaton vacuum case $b=0$. The
question of assigning a temperature to such massless black holes
might be evacuated by arguing that they cannot be formed, either
through central collapse of matter, or (as we have seen above) through
evaporation of massive black holes. Nevertheless, as a matter of principle
one should consider the possibility of primordial massless black
holes. From the general temperature law (\ref{th}) these should have a
finite temperature. On the other hand, being massless they cannot
radiate energy away, so their temperature should vanish.

The question can be settled by solving the massless Klein-Gordon
equation in the metric (\ref{nbhm0}) with $b=0$,
\begin{equation}
\label{cm}
ds^2 = \frac{r}{r_0}dt^2 - \frac{r_0}{r}dr^2 - r_0rd\Omega^2 \,.
\end{equation}
This metric can be rewritten as
\begin{equation}
ds^2 = \Sigma^2\bigg[d\tau^2 - dx^2 - d\Omega^2\bigg] \,,
\end{equation}
with
\be
x = \ln(r/r_0)\,, \quad \tau = t/r_0\,, \quad \Sigma = r_0\e^{x/2}\,,
\ee
showing that the linear dilaton vacuum metric is conformal
to the product $M_2 \times S_2$ of a two-dimensional Minkowski
spacetime with the two-sphere.
Performing also the redefinition
\begin{equation}
\phi = \Sigma^{-1}\psi \,,
\end{equation}
the Klein-Gordon equation (\ref{kg}) is reduced to
\be
\nabla^2\phi = \Sigma^{-3}\bigg[\partial_{\tau\tau} - \partial_{xx} -
\nabla_{\Omega}^2 + \frac14\bigg]\psi = 0\,,
\ee
where $\nabla_{\Omega}^2$ is the Laplacian operator on the
two-sphere.

For a given spherical harmonic with orbital quantum number
$l$, the reduced Klein-Gordon equation is thus
\be
\nabla_2^2\psi_l + (l+1/2)^2\psi_l = 0\,,
\ee
with $\nabla_2^2$ the Dalembertian operator on $M_2$.
Also, for a given spherical harmonic the four-dimensional
Klein-Gordon norm reduces to the $M_2$ norm:
\be
\Vert\phi\Vert^2 = \frac1{2i}\int
d^3x\sqrt{|g|}g^{0\mu}\phi^{*}\stackrel{\leftrightarrow}{\partial_{\mu}}\phi
= \frac{2\pi}{i}\int
dx\,\psi_l^{*}\stackrel{\leftrightarrow}{\partial_{\tau}}\psi_l\,.
\ee
Thus, the problem of wave propagation in the linear dilaton vacuum
reduces to the propagation of eigenmodes of a free Klein-Gordon field
in two dimensions, with effective mass $\mu = l+1/2$. Clearly there is
no reflection, so that the linear dilaton vacuum does not radiate and
hence its Hawking temperature vanishes, contrary to the naive surface
gravity value (\ref{th}). A similar reasoning holds in 2+1 dimensions
for the BTZ vacuum \cite{BTZ} ($M=L=0$), which is conformal to $M_2
\times S_1$.

We have shown that a complete analytical computation of the radiation
spectrum is possible for linear dilaton black hole solutions of
EMD. For massive black holes, this leads in the high frequency regime
to a Planckian distribution with a temperature independent of the
black hole mass, in accordance with the surface gravity value. On the
other hand, we find that extreme, massless black holes do not radiate,
thereby solving the paradox presented by apparently hot (if the
surface gravity temperature is taken seriously) yet massless black holes.
\par
\noindent {\bf Acknowledgements:} J.C.F. thanks the LAPTH for the
warm hospitality during the elaboration of this work. He also thanks
CNPq (Brazil) for partial support. J.C.F. and G.T.M. thank the
French-Brazilian scientific cooperation CAPES/COFECUB for partial
financial support.


\begin{thebibliography}{20}

\bibitem{birrell} N.D. Birrell and P.C.W. Davies, {\bf Quantum fields
in curved space}, Cambridge University Press, Cambridge (1982).

\bb{haw} S.W. Hawking, Commun. Math. Phys. {\bf 43} (1975) 199.

\bb{unruh1} W.G. Unruh, Phys. Rev. D{\bf 14} (1976) 870.

\bb{page} D. Page, Phys. Rev. D{\bf 13} (1976) 198.

\bb{unruh2} W.G. Unruh, Phys. Rev. D{\bf 14} (1976) 3251.

\bb{BTZ} M. Ba\~{n}ados, C. Teitelboim and J. Zanelli, Phys. Rev.
Lett. {\bf 69} (1992) 1849.

\bb{GL} K. Ghoroku and A.L. Larsen, Phys. Lett. B{\bf 328} (1994) 28.

\bb{NSS} M. Natsuume, N. Sakai and M. Sato, Mod. Phys. Lett. A{\bf
11} (1996) 1467.

\bb{BSS} D. Birmingham, I. Sachs and S. Sen, Phys. Lett. B{\bf 413}
(1997) 281.

\bb{CHM} K.C.K. Chan, J.H. Horne and R.B. Mann, Nucl. Phys. B{\bf
447} (1995) 441.

\bibitem{newdil2} G. Cl\'ement and C. Leygnac, Phys. Rev. D{\bf
70} (2004) 084018.

\bb{GM} G.W. Gibbons and K. Maeda, Nucl. Phys. B{\bf 298} (1988)
741.

\bb{GHS} D. Garfinkle, G.T. Horowitz and A. Strominger, Phys. Rev.
D{\bf 43} (1991) 3140.

\bb{hayward} S.A. Hayward, Class. Quantum Grav. {\bf 17} (2000) 4021.

\bb{HHR} S.W. Hawking, G.T. Horowitz and S.F. Ross, Phys. Rev. D{\bf
51} (1995) 4302.

\bb{newdil1} G. Cl\'ement, D. Gal'tsov and C. Leygnac, Phys. Rev.
D{\bf 67} (2003) 024012.

\end{thebibliography}
\end{document}